\pdfoutput=1
\documentclass[11pt,twoside,a4paper,cmspaper,final,collab]{cms-tdr}
\def\svnVersion{386a71d-D}\def\svnDate{2021/09/01}\def\cmsCernNoTag{CERN-EP-2021-039}\def\cmsCernDate{\today}\def\cmsMessage{"Published in Physical Review Letters as \href{http://dx.doi.org/10.1103/PhysRevLett.127.102002}{\doi{10.1103/PhysRevLett.127.102002}.}"}
\begin{document}\cmsNoteHeader{HIN-19-003}

\newlength\cmsTabSkip\setlength{\cmsTabSkip}{1ex}

\newcommand {\RAA}     {\ensuremath{R_{\mathrm{AA}}}\xspace}
\newcommand {\TAA}     {\ensuremath{T_{\mathrm{AA}}}\xspace}
\newcommand {\NAA}     {\ensuremath{N_{\mathrm{AA}}}\xspace}
\newcommand {\mll}     {\ensuremath{m_{\ell\ell}}\xspace}
\newcommand {\vTwo}      {\ensuremath{v_{2}}\xspace}
\newcommand {\ptz}     {\ensuremath{\pt^{\PZ}}\xspace}
\newcommand {\ptl}     {\ensuremath{\pt^{\ell}}\xspace}
\newcommand {\yz}      {\ensuremath{y_{\PZ}}\xspace}
\newcommand {\phiz}    {\ensuremath{\phi_{\PZ}}\xspace}
\newcommand {\PsiTwo}    {\ensuremath{\Psi_2}\xspace}
\newcommand {\lplm}    {\ensuremath{\ell^{+}\ell^{-}}\xspace}
\newcommand {\etaL}    {\ensuremath{\eta^{\ell}}\xspace}
\newcommand {\Aphi}    {\ensuremath{A_{\phi}}\xspace}
\newcommand {\sigmaNNZ}{\ensuremath{\sigma_{\mathrm{NN}}^{\PZ}}\xspace}
\newcommand {\sigmaPP} {\ensuremath{\sigma_{\Pp\Pp}}\xspace}

\newcommand {\pp}      {\ensuremath{\Pp\Pp}\xspace}
\newcommand {\pPb}     {\ensuremath{\Pp}Pb\xspace}
\newcommand {\PbPb}    {\ensuremath{\text{PbPb}}\xspace}
\ifthenelse{\boolean{cms@external}}{\providecommand{\suppMaterial}{the supplemental material~\cite{supp}}}{\providecommand{\suppMaterial}{Appendix~\ref{app:suppMat}\nocite{supp}\xspace}}

\cmsNoteHeader{HIN-19-003}

\title{Constraints on the initial state of PbPb collisions via measurements of \texorpdfstring{$\PZ$}{Z} boson yields and azimuthal anisotropy at \texorpdfstring{$\sqrtsNN = 5.02\TeV$}{sqrt(sNN) =  5.02 TeV}}

\date{\today}

\abstract{
The CMS experiment at the LHC has measured the differential cross sections of \PZ bosons decaying to pairs of leptons, as functions of transverse momentum and rapidity, in lead-lead collisions at a nucleon-nucleon center-of-mass energy of 5.02\TeV.  The measured \PZ boson elliptic azimuthal anisotropy coefficient is compatible with zero, showing that \PZ bosons do not experience significant final-state interactions in the medium produced in the collision.  Yields of \PZ bosons are compared to Glauber model predictions and are found to deviate from these expectations in peripheral collisions, indicating the presence of initial collision geometry and centrality selection effects.  The precision of the measurement allows, for the first time, for a data-driven determination of the nucleon-nucleon integrated luminosity as a function of lead-lead centrality, thereby eliminating the need for its estimation based on a Glauber model.}

\hypersetup{
pdfauthor={CMS Collaboration},
pdftitle={Constraints on the initial state of PbPb collisions via measurements of Z boson yields and azimuthal anisotropy at sqrt(sNN) =  5.02 TeV},
pdfsubject={CMS},
pdfkeywords={CMS, Z boson, heavy ions, PbPb, 5.02 TeV, electroweak boson, initial state, npdf}}

\maketitle
 
Ultrarelativistic heavy ion collisions produce a hot partonic medium, known as the quark-gluon plasma (QGP)~\cite{Busza_2018}.  Energetic partons produced in the collision lose energy traversing the QGP, resulting in a phenomenon known as ``jet quenching''~\cite{Bjorken:1982tu,d'Enterria:2009am}. The magnitude of parton energy loss is frequently quantified with the nuclear modification factor (\RAA) defined as the ratio of a given particle's yield in nucleus-nucleus collisions (\NAA) with the corresponding proton-proton cross section (\sigmaPP), scaled to account for the number of binary nucleon-nucleon interactions in the heavy ion collision~\cite{d'Enterria:2009am}.  For a collision impact parameter (or centrality) $b$, $\RAA(b) = \NAA(b)/[\sigmaPP\TAA(b)]$, where \TAA is the transverse overlap function representing the collision's effective integrated nucleon-nucleon luminosity.  Head-on (central) collisions have larger \TAA than glancing (peripheral) collisions.  A Monte Carlo (MC) Glauber model is typically used to determine $\TAA(b)$ from transverse profiles of colliding nuclei~\cite{Miller:2007ri}.  If the nucleus-nucleus collision is a superposition of independent nucleon-nucleon collisions, \RAA is unity.  Deviations from unity typically indicate the presence of initial- and/or final-state effects.

Since \PZ bosons and their leptonic daughters carry no color charge, they are unaffected by final-state QGP effects and provide a clean test of both \TAA Glauber scaling and modifications of the nuclear parton distribution functions (nPDFs) compared to the free proton case~\cite{Aad:2012ew,Chatrchyan:2014csa,Aad:2019lan,Acharya:2020puh}.  Because of their large mass and clean final state, \PZ boson yields can be both precisely measured and accurately calculated using perturbative quantum chromodynamics (QCD).  Previous measurements by the ALICE, ATLAS, and CMS Collaborations found \PZ boson yields to be uniform in azimuth and consistent with \pp cross sections scaled by Glauber model expectations~\cite{Aad:2012ew,Chatrchyan:2014csa,Aad:2019lan,Acharya:2020puh}.  Measurements involving \PW bosons~\cite{Aad:2019sfe,Chatrchyan:2012nt} and photons~\cite{Adam:2015lda,Aad:2015lcb,Chatrchyan:2012vq} give similar results.  However, the precision of these studies was statistically limited for peripheral collisions. Measurements of \RAA for colored hard probes such as high transverse momentum (\pt) hadrons, jets, and quarkonia, showed a suppression, with respect to Glauber scaling expectations, in peripheral events where limited QGP production is expected, presenting a challenge to theoretical interpretations~\cite{Arleo:2017ntr}. In this region, the Glauber model has large uncertainties from nucleon fluctuations and other sub-nucleon and nuclear structure effects~\cite{dEnterria:2020dwq}. Thus, understanding the onset of jet quenching as a function of centrality remains a key open question. Furthermore, the observation of QGP-like phenomena in the small systems produced in \pp and proton-lead (\pPb) collisions~\cite{Abelev:2012ola,Aad:2012gla,Khachatryan:2016txc,Dusling:2015gta,Nagle:2018nvi} indicates the possibility of final-state effects in such systems.  A precision measurement of \PZ bosons in peripheral nucleus-nucleus collisions can provide an experimental reference for the expected yields of hard probes in the absence of final-state effects, which may lead to an improved understanding of the onset of jet quenching in small systems. 

In this Letter, the yields of \PZ bosons decaying to pairs of muons or electrons are measured using the 2018 data set of \PbPb collisions recorded by the CMS experiment at a nucleon-nucleon center-of-mass energy of $\sqrtsNN = 5.02\TeV$.  Contributions from virtual photons decaying to lepton pairs ($\PGg^*\to\lplm$, where $\ell = \Pgm, \Pe$), over the invariant mass range $60 < \mll < 120\GeV$, are included in the \PZ boson signal.  The results are compared with Glauber model predictions for \TAA scaling of hard, colorless probes.  A measurement of the \PZ boson elliptic azimuthal anisotropy coefficient (\vTwo) is also presented.  The observable \vTwo is defined as the average of $\cos[2(\phiz-\PsiTwo)]$, where \phiz is the \PZ boson azimuthal angle and \PsiTwo is the angle of maximum azimuthal particle density~\cite{Luzum:2012da}.  Finally, the \PZ boson transverse momentum (\ptz) and rapidity (\yz) distributions are measured.

The central feature of the CMS apparatus is a superconducting solenoid of 6\unit{m} internal diameter, providing a magnetic field of 3.8\unit{T}. Within the solenoid volume are a silicon pixel and strip tracker, a lead tungstate crystal electromagnetic calorimeter (ECAL), and a brass and scintillator hadron calorimeter, each composed of a barrel and two endcap sections. Forward hadron (HF) calorimeters extend the pseudorapidity ($\eta$) coverage provided by the barrel and endcap detectors. Muons are detected in gas-ionization chambers embedded in the steel flux-return yoke outside the solenoid.  Events of interest are selected using a two-tiered trigger system~\cite{Sirunyan:2020zal,Khachatryan:2016bia}.  A more detailed description of the CMS detector, together with a definition of the coordinate system used and the relevant kinematic variables, can be found in Ref.~\cite{Chatrchyan:2008zzk}.

The analyzed data corresponds to an integrated luminosity of $1.696 \pm 0.032\nbinv$. Events containing at least one muon (electron) with $\ptl > 12\,(20)\GeV$ are selected by the trigger.  Collision centrality is determined from the total transverse energy deposited in both HF calorimeters. It is expressed as a percentage of the total hadronic cross section, with lower values corresponding to central collisions and larger values to peripheral collisions.  The selected data correspond to $11.78 \pm 0.15$ billion sampled minimum-bias (MB) events in the 0--90\% centrality range.  The \TAA values listed in \suppMaterial{} are used to normalize different centrality selections for comparison and are calculated with \textsc{TGlauberMC} v3.2~\cite{Loizides:2017ack}.

{\tolerance=1400 Detector acceptance and response effects are corrected using simulated MC events.  The signal sample consists of $\PZ/\PGg^{*}\to\lplm$ production generated with \MGvATNLO (v2.4.2)~\cite{Alwall:2014hca} at next-to-leading order (NLO) in QCD.  These events are interfaced with ``tuned''~\cite{Sirunyan:2019dfx} \PYTHIA 8 (v2.3.0)~\cite{Sjostrand:2014zea} to simulate parton showering, hadronization, and underlying event production.  Simulations account for the lead nucleus isospin content and use EPPS16 NLO nPDFs~\cite{Eskola:2016oht} combined with free-nucleon CT14 NLO PDFs~\cite{Dulat:2015mca}.  This setup does not account for potential neutron skin effects, but such effects are expected to be small for neutral electroweak bosons~\cite{Helenius:2016dsk}.  The \PZ boson events are then overlaid onto \PbPb events produced with \HYDJET v1.9~\cite{Lokhtin:2005px}.  Those events are then passed to \GEANTfour~\cite{Agostinelli:2002hh} to emulate detector response. \par}
Leptons are reconstructed using the CMS particle flow algorithm~\cite{Sirunyan:2017ulk}.  This algorithm applies loose isolation criteria to muons, but not for electrons.  Muon (electron) candidates must have $\abs{\etaL} < 2.4\,(2.1)$ and $\ptl > 20\GeV$.  Selection criteria are required to reject low-quality muons~\cite{Sirunyan:2018fpa}, resulting in $\approx$98\% identification efficiency.  The effects of misreconstructed muons are negligible.  A multivariate discriminant optimized using the \textsc{TMVA} package~\cite{Hocker:2007ht} selects electrons~\cite{Khachatryan:2015hwa} with a working point corresponding to 90\% identification efficiency and 80\% rejection of misreconstructed electrons.  Efficiencies of various stages of lepton reconstruction are measured using the ``tag-and-probe'' method~\cite{Khachatryan:2010xn} in data and simulation, as functions of \ptl, \etaL, and centrality.  This includes lepton trigger efficiency, the efficiency of reconstructing a muon (electron) track that matches a muon detector plane (ECAL tower), and the probability that the lepton passes all selections.  Discrepancies between data and simulation are corrected by weighting the simulation lepton efficiency by the ratio of the data and MC efficiencies.

The \PZ boson candidates are reconstructed from oppositely charged lepton pairs with $60 < \mll < 120\GeV$ and $\abs{\yz} < 2.1$.  This results in 19\,104 (9863) candidates in the dimuon (dielectron) channel. The \PZ boson reconstruction efficiency ($\epsilon$) is calculated using the MC sample as functions of \ptz, \yz, and centrality.  The efficiency is around 70--95 (30--65)\% from central to peripheral events in the dimuon (dielectron) channel.  Candidates are weighted by $1/\epsilon$ to correct for lepton selection and detector inefficiencies.  A similar correction for detector acceptance is applied to account for \PZ bosons produced within $\abs{\yz} < 2.1$ but having decays outside the \ptl or \etaL selections. The average acceptance is 0.68 in the dimuon channel, but only 0.58 in the dielectron channel because of the smaller \etaL range allowed.

Multiple background sources can create high-mass lepton pairs.  The first is from QCD-initiated hard processes, such as the production of two leptons inside jets.  Because this background arises largely from random lepton combinations, it is assumed that the production rates of same-sign and opposite-sign lepton pairs are equal.  A total of 44 (167) same-sign lepton pairs are observed in the dimuon (dielectron) channel.  In the electron channel, misreconstruction of the electron charge slightly enhances the same-sign yield.  After correcting for this effect, this background is 0.2 (1.0)\% of the opposite-sign yield in the dimuon (dielectron) channel.  A second background is generated by electromagnetic (EM) processes (\eg, $\PGg\PGg\to\lplm$)~\cite{CMS:2020avp}.  Here, the photons are emitted by the incoming nuclei and tend to have very low \pt~\cite{Sirunyan:2018fhl}.  Thus, the lepton pair \pt strongly peaks near zero, and the daughter leptons are back-to-back in azimuth ($\phi$).  Based on simulated \textsc{starlight} v2.2~\cite{Klein:2016yzr} events, any dimuon (dielectron) candidates having $\pt < 1.25\,(2.50)\GeV$ and acoplanarity, defined as $\Aphi = 1-\Delta\phi/\pi$, less than 0.001 are identified as products of EM background.  The \pt threshold for the dielectron channel is larger because of the worse energy resolution of electrons compared to muons.  In simulated events these selections correspond to 90\% background rejection, and result in a small efficiency loss for \PZ bosons, which is taken care of with the applied corrections.  Candidates resulting from this background account for 0.6 (0.7)\% of the dimuon (dielectron) yield before subtraction.  The other backgrounds considered are $\PZ\to\Pgt^{+}\Pgt^{-}$, $\ttbar$ production, and the production of \PW bosons decaying to a lepton that is combined with another lepton originating from a hadron decay.  The expected yields are calculated as functions of centrality, \ptz, or \yz using appropriate MC samples.  These backgrounds contribute less than 0.3\% to the total yield in each channel. 

The \ptz resolution is around 6.5 (7.7)\% in the dimuon (dielectron) channel.  When measuring the \ptz spectrum, this results in the migration of \PZ candidates between bins.  This is corrected using an unregularized matrix inversion unfolding procedure implemented with the \textsc{RooUnfold} framework~\cite{Adye:2011gm}.  A cross-check using a regularized Bayesian method was found to give consistent results~\cite{DAgostini:1994fjx}.  Systematic uncertainties related to mismodeling of the \ptz distribution's shape are negligible.  However, the statistical uncertainty of the MC response matrix is propagated to the final spectrum as a systematic uncertainty.  This uncertainty is up to 2 (4)\% for the dimuon (dielectron) channel in the lowest \ptz bin, but is $<$1\,(2)\% at higher \ptz.

To measure $\vTwo$, the 3-subevent scalar product method of Refs.~\cite{Luzum:2012da,Sirunyan:2017plt} is used.  This technique compares $\phiz$ to the global event azimuthal shape measured using the HF calorimeters and midrapidity charged particles.  

\begin{figure}[]
\centering
  \includegraphics[width=0.49\textwidth]{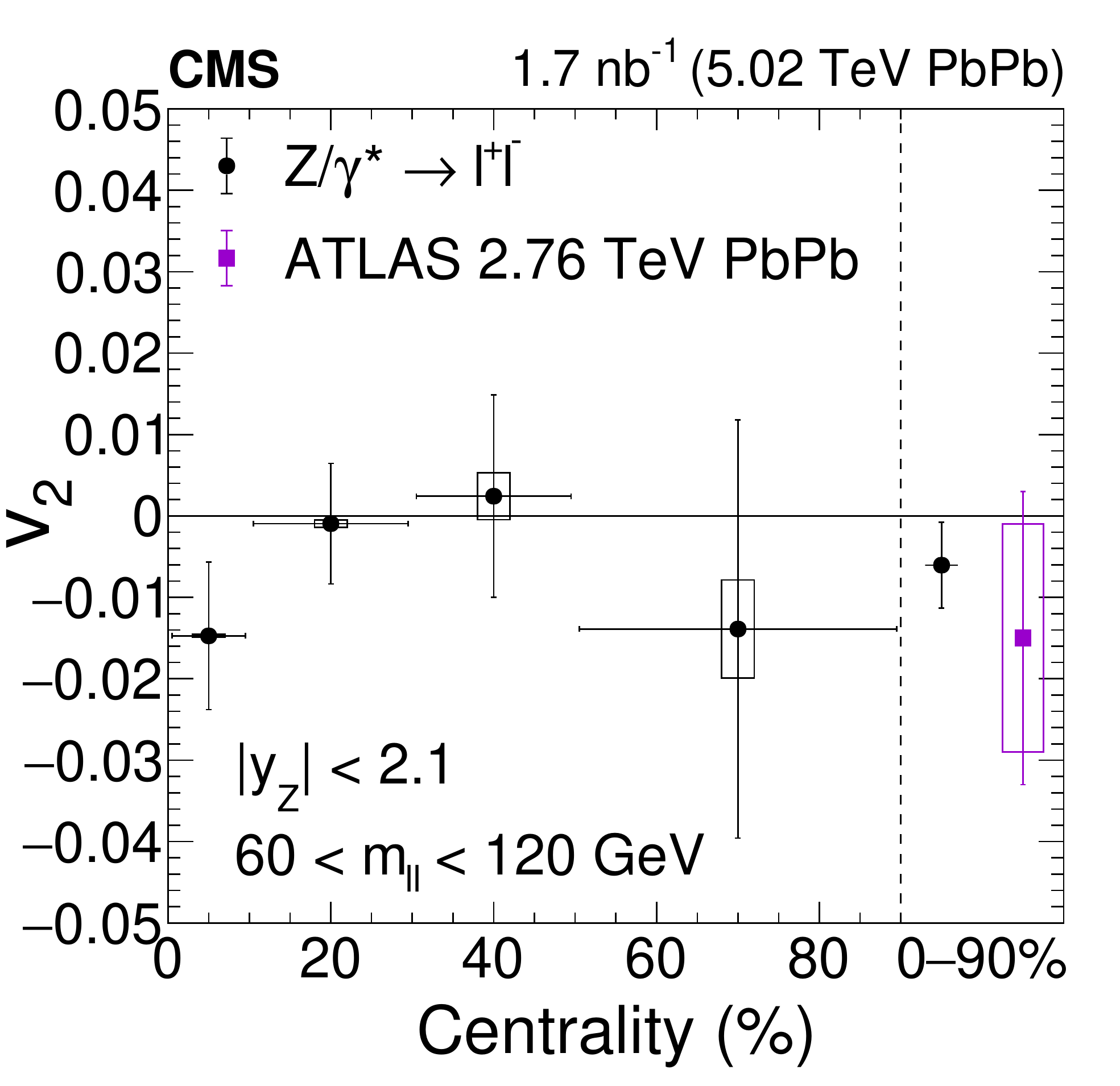}
  \caption{The \vTwo of \PZ bosons in \PbPb collisions for various centrality bins.  The error bars represent statistical uncertainties.  The boxes represent systematic uncertainties and may be smaller than the markers.  A measurement from the ATLAS Collaboration at $\sqrtsNN = 2.76\TeV$ is also shown~\cite{Aad:2012ew}.}
  \label{fig:Zv2}
\end{figure}

The centrality calibration is affected by the MB event selection efficiency of the HF calorimeters, which is $97.5^{+1.0}_{-0.5}$\% for the 0--100\% centrality range. The uncertainty in this efficiency is propagated to the final observables, resulting in a final uncertainty of 0.1 (8.4)\% in central (peripheral) events. Uncertainties in the single-lepton trigger, reconstruction, and selection efficiencies are the dominant sources of uncertainty and are calculated with the tag-and-probe procedure.  After accounting for each \PZ boson decay daughter, this effect propagates into a 3.0 (5.9)\% uncertainty in the cross sections measured in the dimuon (dielectron) channel.  An additional uncertainty of less than 1\% accounts for the statistical uncertainty of the MC sample used to calculate the \PZ boson efficiency.  The model dependence of the acceptance correction is calculated to be 0.6\% by examining the impact of using different nPDF Hessian error sets~\cite{Butterworth:2015oua}. The effect of electron charge misreconstruction is 0.5\% in simulation.  Differences between the electric charge sign-flip probability in data and simulation are estimated to be less than a factor of two, so an absolute uncertainty of 0.5\% is assigned for this effect. The \Aphi and \pt selection criteria for removing EM backgrounds are varied between working points corresponding to 80 and 95\% background rejection to gauge the sensitivity of the analysis to these selections.  This results in an uncertainty of up to 1.5\% in the 70--90\% centrality range but is negligible elsewhere.  When evaluated as a function of \ptz, this uncertainty is 4\% in the lowest \ptz bin, but decreases for higher \ptz values.  Uncertainties related to backgrounds estimated from simulation are negligible. Both decay channels are combined into a single measurement using the BLUE method~\cite{Valassi:2003mu}.  Correlations between channels resulting from the centrality calibration, number of MB events, \TAA, and the acceptance correction are accounted for.  For all observables, both channels are within 1.5 standard deviations ($\sigma$) of each other.

\begin{figure}[]
\centering
    \includegraphics[width=0.49\textwidth]{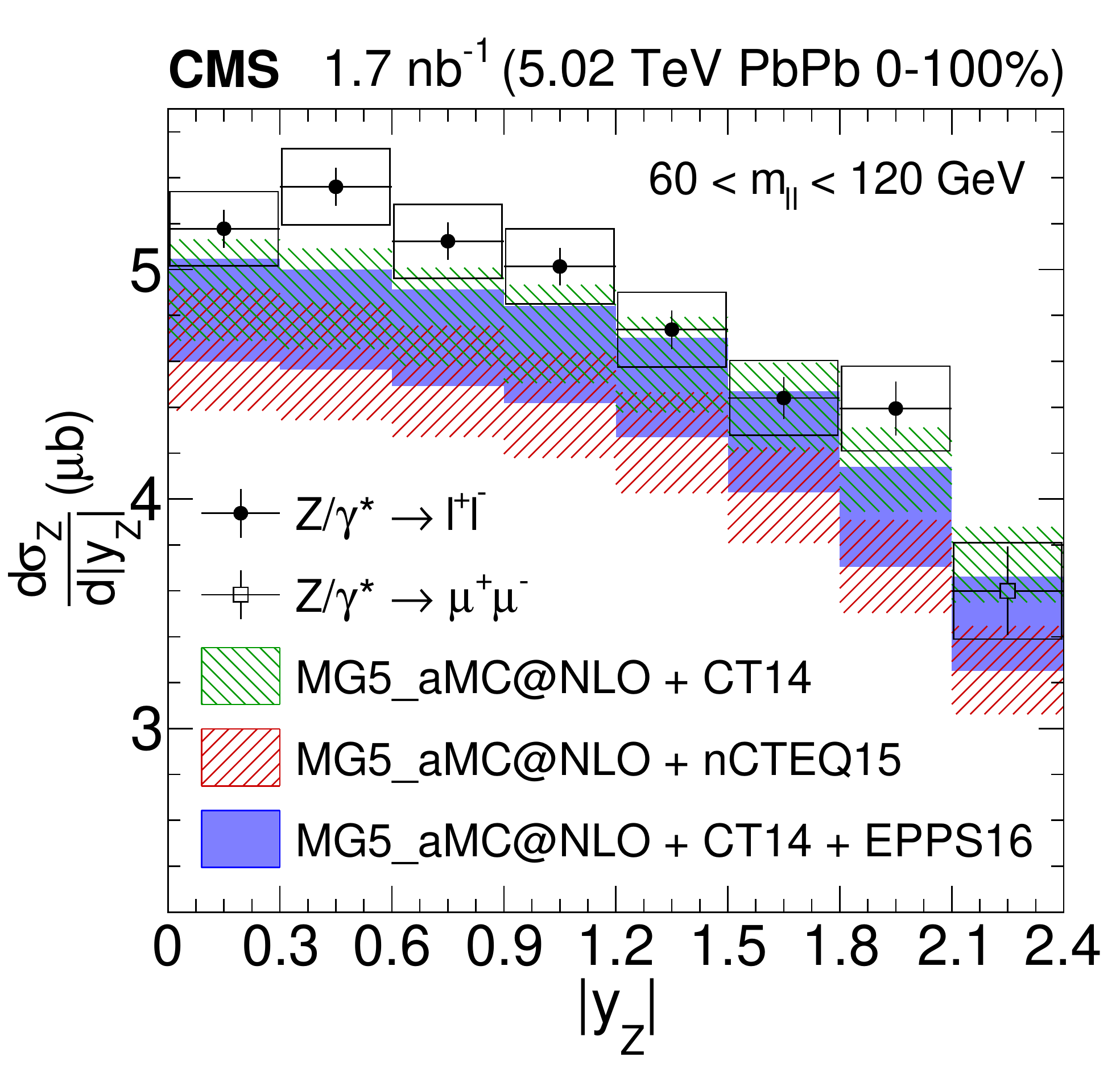}
  \caption{ The \PZ boson differential cross section as a function of $\abs{\yz}$.  The error bars represent statistical uncertainties, while the boxes represent systematic uncertainties.  Predictions using one PDF and two different nPDF sets are also shown.}
  \label{fig:ZYields_y}
\end{figure}
The \PZ boson \vTwo as a function of centrality is shown in Fig.~\ref{fig:Zv2}.  A previous measurement from the ATLAS Collaboration at $\sqrtsNN = 2.76\TeV$ is also shown~\cite{Aad:2012ew}.  The new measurement in the 0--90\% range is compatible with zero and is significantly more precise than the previous measurement.  This confirms the expectation that \PZ bosons, being produced and decaying very early in the collision process, are largely unaffected by final-state effects such as hydrodynamic flow and energy loss.

\begin{figure}[]
\centering
    \includegraphics[width=0.49\textwidth]{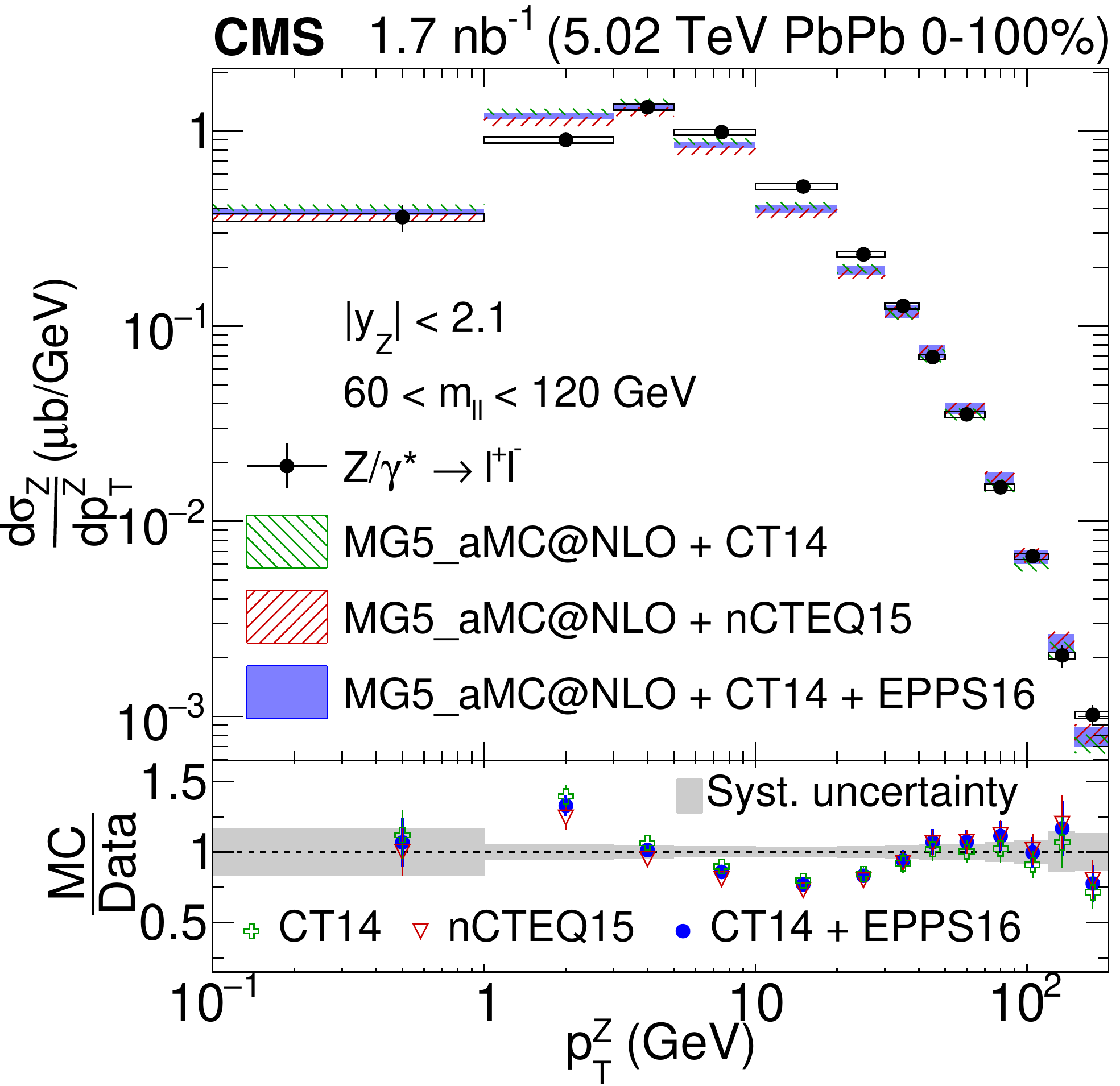}
  \caption{ The \PZ boson differential cross section as a function of \ptz.  The error bars represent statistical uncertainties, while the boxes represent systematic uncertainties.  Predictions using one PDF and two different nPDF sets are also shown.  The lower panel shows the ratio of the predictions to data.}
  \label{fig:ZYields_pt}
\end{figure}

The differential cross section of \PZ boson production as a function of $\abs{\yz}$ for 0--100\% centrality events is shown in Fig.~\ref{fig:ZYields_y}.  Predictions from the \MGvATNLO MC generator interfaced with the CT14 free proton PDF~\cite{Dulat:2015mca}, as well as the CT14+EPPS16~\cite{Eskola:2016oht} and nCTEQ15~\cite{Kovarik:2015cma} nPDF sets are also shown.  The models predictions are scaled up by the square of the atomic mass number $A^2$, where $A = 208$ for lead.  The width of the model band indicates the PDF and QCD scale uncertainties added in quadrature.  The data lie on the upper edge of the models.  Inclusion of missing higher-order (next-to-NLO) corrections in the theoretical predictions would increase the total \PZ boson cross section by a few percent~\cite{Sirunyan:2019crt}.  The predictions derived with the EPPS16 nPDF are closer to the data than those computed with the nCTEQ15 set.  Differential cross sections as a function of \ptz are shown in Fig.~\ref{fig:ZYields_pt}.  The \ptz distribution peaks around 5\GeV before falling sharply.  The spectrum is compared to the \MGvATNLO calculations with different (n)PDF sets.  Although the general trend of the data is correctly reproduced by the simulation, some discrepancies are apparent in different \ptz regions.  At low \ptz, this could be related to the lack of soft gluon resummation in the model, as discrepancies have also been observed in \pp collisions~\cite{Sirunyan:2019bzr}.  The nPDF sets are around 10\% lower than the free proton PDF at low \ptz, but this trend is reversed for $\ptz > 50\GeV$.  However, the differences between the two nPDF sets are smaller than the deviation of the models from the data, indicating that modeling improvements are needed to reproduce the measured \ptz spectrum before considering nPDF modifications. 

\begin{figure}[t]
\centering
    \includegraphics[width=0.49\textwidth]{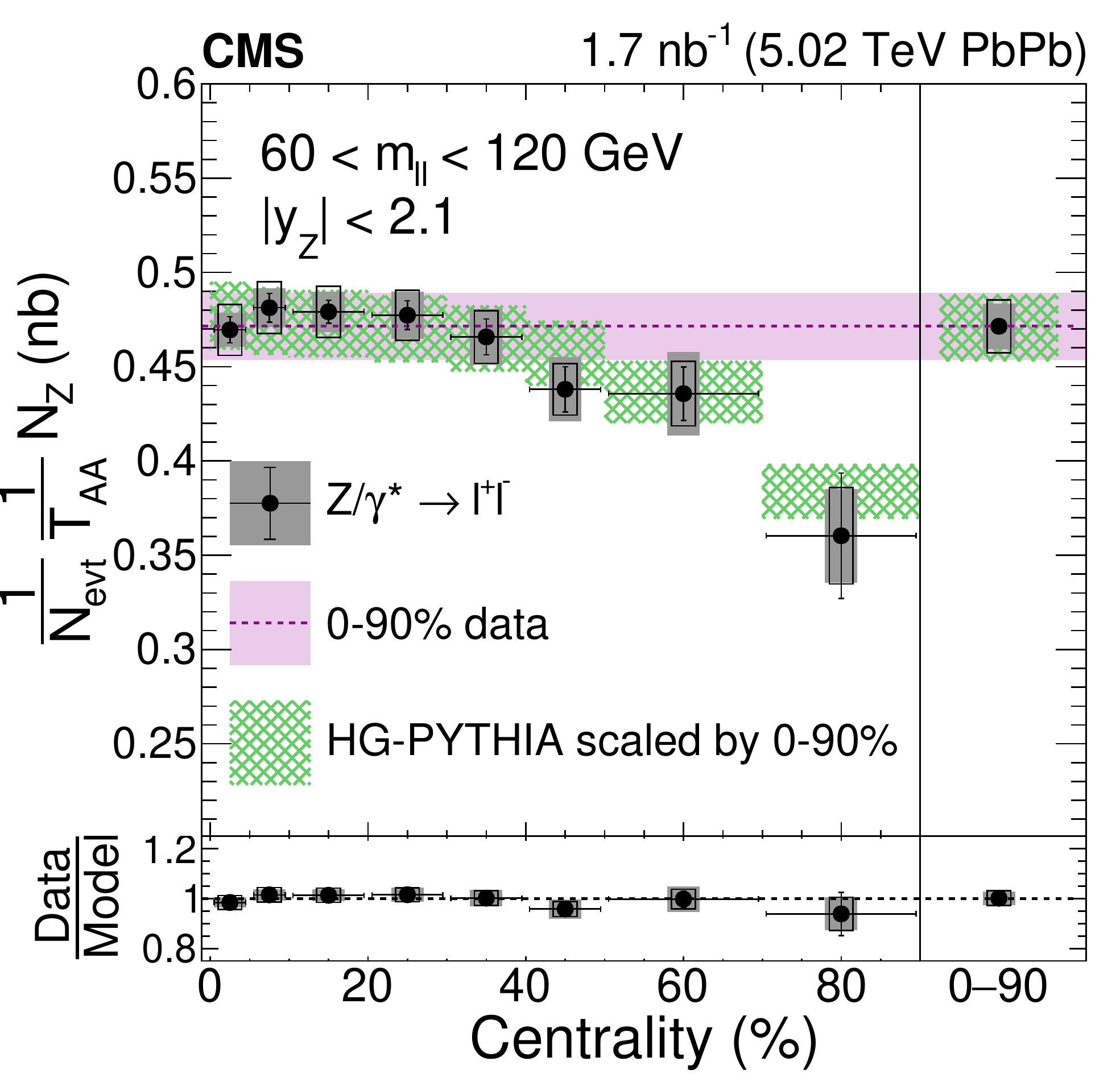}
  \caption{The \TAA-normalized yields of \PZ bosons as a function of centrality.  The error bars, hollow boxes, and solid gray boxes represent the statistical, systematic, and \TAA uncertainties, respectively.  The value of the 0--90\% data point, and the scaled \textsc{HG-PYTHIA} model are shown for comparison, with the width of the bands representing the contribution from the total 0--90\% data point uncertainty.}
  \label{fig:ZYields}
\end{figure}

The \TAA-normalized \PZ boson yields are shown in Fig.~\ref{fig:ZYields} versus centrality.  Previous analyses of \PZ boson production~\cite{Chatrchyan:2014csa,Aad:2012ew,Aad:2019lan} have indicated that the \TAA-scaling assumption works well in central events and also for inclusive selections.  Comparison of the data for centralities less than 40\% with the inclusive 0--90\% data point, shown by the dashed magenta line, confirms these previous results.  However, a decreasing trend for the 40--90\% centrality range can be seen, which was not observed with less precise measurements~\cite{Chatrchyan:2014csa,Aad:2012ew}.  The significance of this deviation from the 0--90\% data point is 1.6$\sigma$ (2.2$\sigma$) in the 40--90 (70--90)\% range.   This depletion is not expected to be caused by any final-state effect, because it would show up more prominently in central events.  

One potential explanation for this trend is the presence of initial-state geometry and centrality selection effects in peripheral events~\cite{Morsch:2017brb}.  Initial-state geometry effects can arise because bound nucleons have a spatial distribution that can shift the average nucleon-nucleon impact parameter away from that of \pp collisions.  Likewise, correlations between the hard process and the soft-particle production yield, which is used for estimating centrality, can shift the average \TAA in a given centrality range away from the Glauber expectation.  Such effects have been studied for high-\pt charged-hadron production by the ALICE and CMS Collaborations~\cite{Sirunyan:2018eqi,Acharya:2018njl}.  The \textsc{HG-PYTHIA} model~\cite{Morsch:2017brb} attempts to describe these geometric and selection effects by using the \HIJING (v1.35)~\cite{Gyulassy:1994ew} event generator to simulate the initial geometry of a heavy ion event.  In the \textsc{HG-PYTHIA} model, particle production is modeled by overlaying a \PYTHIA 8 (v2.4.3)~\cite{Sjostrand:2014zea} event for each nucleon-nucleon scattering. These ``heavy ion'' events contain no QGP-related physics.  To compare this model to the data, the \RAA for hard scatterings was calculated using the model, before being scaled by the 0--90\% data point to set the overall normalization. The \textsc{HG-PYTHIA} expectation is plotted as the hatched green boxes in Fig.~\ref{fig:ZYields}, which are consistent with the data, indicating that the downward trend in peripheral events can be largely explained by a combination of geometric and selection effects. This result does not support the apparent increasing trend of the \TAA-scaled \PZ boson yields measured in peripheral \PbPb collisions by the ATLAS experiment~\cite{Aad:2019lan}, which was interpreted as an indication of a possible ``shadowed'' value of the nucleon-nucleon inelastic cross section in Ref.~\cite{Eskola:2020lee}.

As the experimental uncertainties are comparable to the Glauber model uncertainties, the quantity $N_{\PZ}/(\sigmaNNZ N_{\text{evt}})$ provides a new experimental proxy for \TAA.  Here, \sigmaNNZ is the \PZ boson production cross section in a nucleon-nucleon collision, which can be estimated using precise \pp measurements~\cite{Aaboud:2018nic}.  This proxy would eliminate the need for Glauber modeling and related assumptions about nuclear structure.  Furthermore, such a quantity would have centrality and selection effects incorporated, thereby allowing for the cancellation of these effects when measuring quantities such as \RAA.  Such cancellations are crucial for studies of the onset of jet quenching in peripheral heavy ion collisions and other small systems, such as \pp and \pPb.

In summary, \PZ boson yields and the elliptic flow coefficient (\vTwo) have been measured with high precision as functions of centrality in lead-lead collisions at a nucleon-nucleon center-of-mass energy of 5.02\TeV.  The \PZ boson \vTwo is compatible with zero, consistent with the expectation of no significant final-state interactions in the quark-gluon plasma (QGP). The differential cross section of \PZ boson production as a function of rapidity is found to lie on the upper edge of theoretical next-to-leading order (NLO) predictions derived with two different nuclear parton distribution functions.  Discrepancies between data and NLO calculations are also observed in various regions of \PZ boson transverse momentum, indicative of missing higher-order theoretical corrections.   Appropriately scaled \PZ boson yields are constant versus impact parameter for central and semi-central collisions, but a decreasing trend is seen for the first time for more peripheral events.  This is compatible with the \textsc{HG-PYTHIA} model prediction, which accounts for initial collision geometry and centrality selection effects.  These results provide a new experimental proxy for estimating the average nucleon-nucleon integrated luminosity as a function of centrality in heavy ion collisions. The ratio of \PZ boson yields in \PbPb over \pp collisions can be used as an alternative to Glauber-model-based scaling for hard scattering processes, which also automatically accounts for potential effects related to event selection and centrality calibration. Such a method provides a useful tool for future searches for the onset of QGP effects on colored hard probes in the medium produced in peripheral heavy ion collisions and small colliding systems.

\begin{acknowledgments}
  We congratulate our colleagues in the CERN accelerator departments for the excellent performance of the LHC and thank the technical and administrative staffs at CERN and at other CMS institutes for their contributions to the success of the CMS effort. In addition, we gratefully acknowledge the computing centers and personnel of the Worldwide LHC Computing Grid and other centers for delivering so effectively the computing infrastructure essential to our analyses. Finally, we acknowledge the enduring support for the construction and operation of the LHC, the CMS detector, and the supporting computing infrastructure provided by the following funding agencies: BMBWF and FWF (Austria); FNRS and FWO (Belgium); CNPq, CAPES, FAPERJ, FAPERGS, and FAPESP (Brazil); MES (Bulgaria); CERN; CAS, MoST, and NSFC (China); MINCIENCIAS (Colombia); MSES and CSF (Croatia); RIF (Cyprus); SENESCYT (Ecuador); MoER, ERC PUT and ERDF (Estonia); Academy of Finland, MEC, and HIP (Finland); CEA and CNRS/IN2P3 (France); BMBF, DFG, and HGF (Germany); GSRT (Greece); NKFIA (Hungary); DAE and DST (India); IPM (Iran); SFI (Ireland); INFN (Italy); MSIP and NRF (Republic of Korea); MES (Latvia); LAS (Lithuania); MOE and UM (Malaysia); BUAP, CINVESTAV, CONACYT, LNS, SEP, and UASLP-FAI (Mexico); MOS (Montenegro); MBIE (New Zealand); PAEC (Pakistan); MSHE and NSC (Poland); FCT (Portugal); JINR (Dubna); MON, RosAtom, RAS, RFBR, and NRC KI (Russia); MESTD (Serbia); SEIDI, CPAN, PCTI, and FEDER (Spain); MOSTR (Sri Lanka); Swiss Funding Agencies (Switzerland); MST (Taipei); ThEPCenter, IPST, STAR, and NSTDA (Thailand); TUBITAK and TAEK (Turkey); NASU (Ukraine); STFC (United Kingdom); DOE and NSF (USA).\end{acknowledgments}

\bibliography{auto_generated}

\ifthenelse{\boolean{cms@external}}{}{
\clearpage

\appendix
\numberwithin{table}{section}
\section{Supplemental material\label{app:suppMat}}
\begin{table}[h]
\centering
\topcaption{The average \TAA values for $\sqrtsNN = 5.02\TeV$ \PbPb collisions.  The uncertainties result from the model parameters and the HF detector energy resolution.}
\begin{scotch}{cc}
Centrality & $\langle \TAA \rangle$ ($\unit{mb}^{-1}$)   \\
 \hline
0--5\% & $25.70\pm0.47$      \\
5--10\% & $20.40\pm0.40$     \\
10--20\% & $14.39\pm0.30$    \\
20--30\% & $8.80\pm0.22$     \\
30--40\% & $5.12\pm0.16$     \\
40--50\% & $2.78\pm0.11$     \\
50--70\% & $0.996\pm0.050$   \\
70--90\% & $0.1650\pm0.0077$ \\
0--90\% & $6.27\pm0.14$      \\
\end{scotch}
\label{tab:TAA}
\end{table}

}

\cleardoublepage \section{The CMS Collaboration \label{app:collab}}\begin{sloppypar}\hyphenpenalty=5000\widowpenalty=500\clubpenalty=5000\vskip\cmsinstskip
\textbf{Yerevan Physics Institute, Yerevan, Armenia}\\*[0pt]
A.M.~Sirunyan$^{\textrm{\dag}}$, A.~Tumasyan
\vskip\cmsinstskip
\textbf{Institut f\"{u}r Hochenergiephysik, Wien, Austria}\\*[0pt]
W.~Adam, F.~Ambrogi, T.~Bergauer, M.~Dragicevic, J.~Er\"{o}, A.~Escalante~Del~Valle, R.~Fr\"{u}hwirth\cmsAuthorMark{1}, M.~Jeitler\cmsAuthorMark{1}, N.~Krammer, L.~Lechner, D.~Liko, T.~Madlener, I.~Mikulec, F.M.~Pitters, N.~Rad, J.~Schieck\cmsAuthorMark{1}, R.~Sch\"{o}fbeck, M.~Spanring, S.~Templ, W.~Waltenberger, C.-E.~Wulz\cmsAuthorMark{1}, M.~Zarucki
% ...
\vskip\cmsinstskip
\dag: Deceased\\
1:  Also at Vienna University of Technology, Vienna, Austria\\
% ...
\end{sloppypar}
%%% END EDITABLE REGION %%%
% skeleton_end
\end{document}